\documentclass[twocolumn,showpacs,amsmath,aps,prl,amssymb,nofootinbib,superscriptaddress]{revtex4-1} 
\usepackage[T1]{fontenc}
\usepackage[latin9]{inputenc}
\usepackage{times}
\usepackage{color} 
\usepackage{xspace}
\usepackage{amssymb,amsmath}
\usepackage{amsbsy}
   \usepackage{dsfont}
\usepackage{graphicx}
\usepackage{bm}
\usepackage{epstopdf}
\usepackage{float}
\usepackage{adjustbox}

\usepackage{soul}

\usepackage[unicode,breaklinks]{hyperref}
\hypersetup{
    unicode=true,
    a4paper=true,
    plainpages=false, 
    colorlinks=true,
    linkcolor=blue,
    citecolor=blue,
    filecolor=black,
    urlcolor=blue
}
\newcommand{\beginsupplement}{%
         \setcounter{table}{0}
         \renewcommand{\thetable}{S\arabic{table}}%
         \setcounter{figure}{0}
         \renewcommand{\thefigure}{S\arabic{figure}}%
          \setcounter{equation}{0}
          \renewcommand{\theequation}{S\arabic{equation}}

      }

\begin{document}
 \title{Quantum droplets of dipolar mixtures}
\author{R. N. Bisset}
\affiliation{Institut f\"ur Theoretische Physik, Leibniz Universit\"at Hannover, Germany}
\affiliation{Institut f\"ur Experimentalphysik, Universit\"at Innsbruck, Innsbruck, Austria}
\author{L. A. Pe\~na Ardila}
\affiliation{Institut f\"ur Theoretische Physik, Leibniz Universit\"at Hannover, Germany}
\author{L. Santos}
\affiliation{Institut f\"ur Theoretische Physik, Leibniz Universit\"at Hannover, Germany}

\begin{abstract}
Recently achieved two-component dipolar Bose-Einstein condensates open exciting possibilities for the study of mixtures of ultradilute quantum liquids. 
While nondipolar self-bound (without external confinement) mixtures are necessarily miscible with an approximately fixed ratio between the two densities, the density ratio for the dipolar case is free.
Therefore, self-bound dipolar mixtures present qualitatively novel and much richer physics, characterized by three possible ground-state phases: 
miscible, symmetric immiscible and asymmetric immiscible, which may in principle occur at any population imbalance. Self-bound immiscible droplets are possible due to mutual nonlocal intercomponent attraction, which results 
in the formation of a droplet molecule. Moreover, our analysis of the impurity regime, shows that quantum fluctuations in the majority component crucially modify the 
miscibility of impurities. Our work opens intriguing perspectives for the exploration of spinor physics in ultradilute liquids, which should resemble to some extent 
that of $^4$He-$^3$He droplets and impurity-doped helium droplets.
\end{abstract} 
  
\maketitle

%%%%%%%%%%%%%%%%%%%%%%%%%%%%%%%%%%%%%%%%%%%%%%%%%%%%%%%%%%%%%%%%%%%%%%%%%%%%%%%%%%%%%%%%%%%

% INTRODUCTION

% Helium droplets and Helium mixtures

\emph{Introduction.--} Helium droplets have been a major focus for many years~\cite{Toennies2001, Toennies2004, Barranco2006, Ancilotto2017}. They remain liquid at low pressures, even at zero temperature, constituting an 
extraordinary scenario for the study of superfluidity down to nanoscopic scales~\cite{Grebenev1998}. Interestingly, helium has two stable isotopes, bosonic $^4$He and fermionic $^3$He, allowing for self-bound droplet mixtures. 
Under a typical experimentally achievable temperature of $0.15$K, $^4$He is a superfluid, whereas $^3$He remains a normal fluid~\cite{Harms1999}. Moreover, due to its smaller mass and limited solubility in $^4$He, $^3$He  resides at the droplet surface surrounding the $^4$He component~\cite{Barranco2006}. Droplets of helium mixtures are hence characteristically phase separated in a core-shell structure, although droplets under rotation may display more intricate distributions~\cite{Pi2020}. Helium droplets can also be doped with other elements or molecules, which may remain at the surface or sink to the core. These crucial properties 
have been extensively explored, both in what concerns the use of embedded dopants to prove superfluidity~\cite{Grebenev1998}, and helium-nanodroplet spectroscopy, i.e.~the use of the pristine low-temperature environment provided by the helium droplet for spectroscopic studies of impurities~\cite{Stienkemeier2006, Choi2006, Tiggesbaeumker2007, Szalewicz2008}. 

% Quantum droplets: binary mixtures and dipolar BEC

Helium droplets constituted up until very recently the only example of a self-bound quantum liquid, confined in the absence of external trapping. 
New developments in the field of ultracold atoms have, however, changed this picture. 
Quantum droplets have been observed both in dipolar Bose-Einstein condensates~(BECs) made of highly magnetic lanthanide 
atoms~\cite{Kadau2016, Chomaz2016, Schmitt2016}, and in binary (nondipolar) homonuclear~\cite{Cabrera2018, Semeghini2018} and heteronuclear~\cite{DErrico2019} Bose mixtures.
Strikingly, these droplets are orders of magnitude more dilute than helium droplets. They are kept self-bound by a mechanism 
known as quantum stabilization~\cite{Petrov2015}:  an almost complete cancellation of the various mean-field forces results in a small residual attraction which is compensated 
by the repulsive Lee-Huang-Yang~(LHY) energy induced by quantum fluctuations. In a dipolar BEC, the mean-field forces are given by the dipolar and contact interactions~\cite{Waechtler2016}, whereas in nondipolar binary mixtures a similar role is played by inter- and intracomponent interactions~\cite{Petrov2015}.

%%%%%%%%%%%%%%%%%%%%%%%%%%%%%%%%% 

% FIGURE 1

\begin{figure} 
\begin{center}
\includegraphics[width=1\columnwidth]{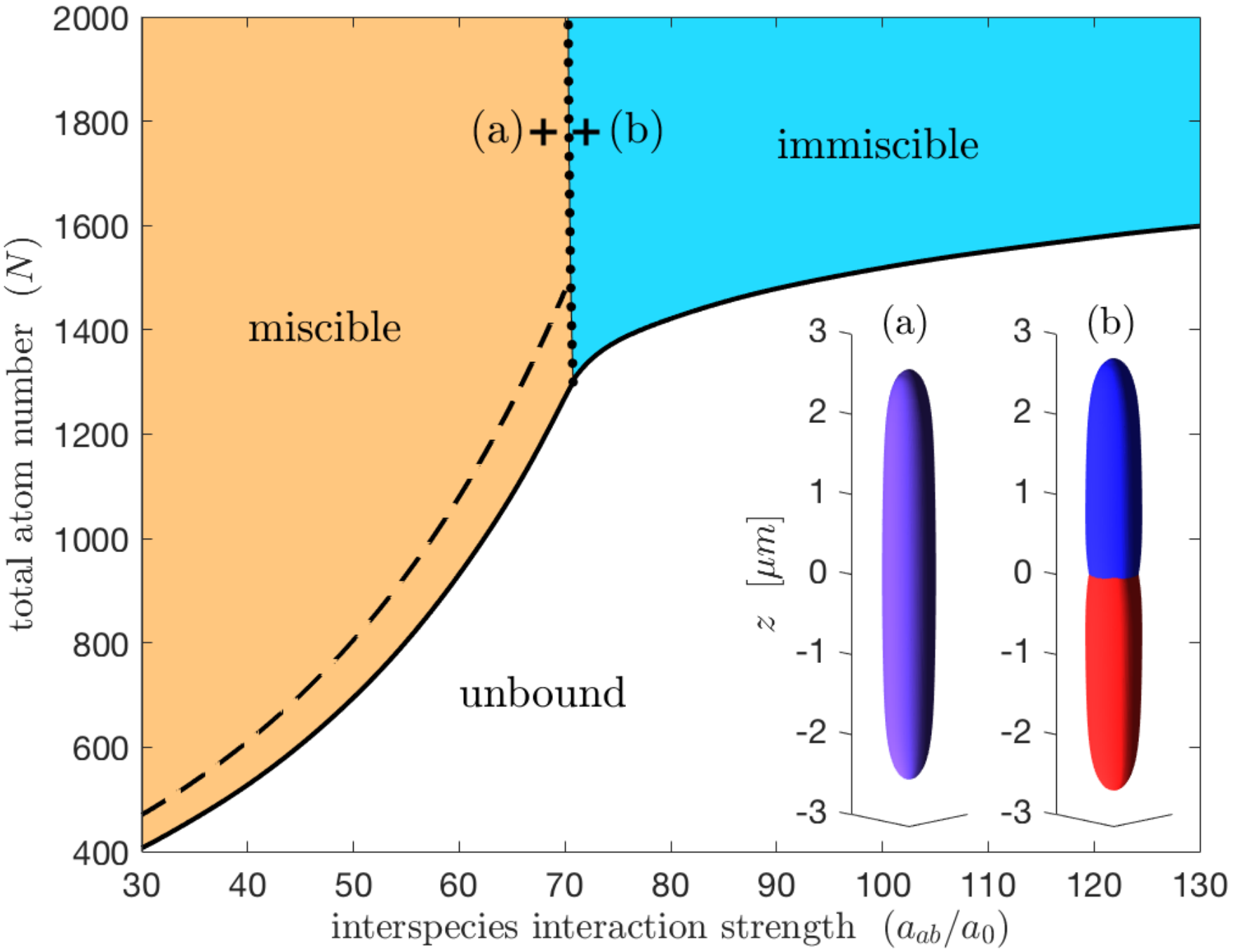}
\caption{Ground-state phase diagram for Dy-Dy mixtures -- in the absence of external confinement -- with $a_{aa}=a_{bb}=70a_0$ and $N_a=N_b=N/2$ as a function of total particle number $N$ and $a_{ab}$. The shaded regions indicate self-bound droplet solutions, whereas below these the solutions are unbound. The dashed curve indicates 
the prediction obtained using the Gaussian Ansatz~\eqref{eq:Gaussian}. The insets show isodensity surface examples for (a) a miscible and (b) an asymmetric immiscible self-bound droplet.}
 \label{fig:3}
\end{center}
\end{figure}

%%%%%%%%%%%%%%%%%%%%%%%%%%%%%%%%% 

The recently observed, ultradilute self-bound mixtures
differ in a crucial way to helium droplet mixtures: they must remain miscible. Moreover, to a good approximation such ultradilute droplets must keep a fixed ratio between the particle number in each component, and deviations from this ratio are evaporated before the droplet sets in. As a result, the spin degree of freedom (i.e.~the 
population difference) remains to a large extent frozen, and the mixture behaves as a single-component BEC~\cite{Petrov2015}. Bose-Fermi mixtures must remain miscible as well~\cite{Rakshit2019}. 

% This Letter

In this Letter, we show that recently realized mixtures of two dipolar species~\cite{Trautmann2018, Durastante2020} open new perspectives for the study of self-bound mixtures 
in which the spin degree of freedom is genuinely free. Self-bound dipolar mixtures may 
be miscible but, crucially, also immiscible~(Fig.~\ref{fig:3}). In the latter scenario, which to the best of our knowledge is unique to dipolar mixtures, the two components phase separate while still being self-bound due to the 
interplay between quantum stabilization and intercomponent dipole-dipole attraction. Moreover, in contrast to experimentally achieved $^3$He-$^4$He droplets,  both components 
should remain superfluid in Bose droplet mixtures under typical experimental conditions. We identify three different ground-state phases for self-bound dipolar mixtures: miscible, symmetric immiscible, and asymmetric immiscible. In contrast to nondipolar mixtures, droplets with any population imbalance (polarization) are possible, all the way from the fully balanced case to the impurity limit~\cite{Wenzel2018}. We show that impurity solubility in a dipolar droplet is crucially affected by quantum fluctuations in the majority component. 
Although we illustrate the possible physics for the case of Dy-Dy mixtures \cite{footnote-atoms}, the qualitative features are generally valid for other dipolar mixtures (in particular Er-Dy~\cite{Trautmann2018, Durastante2020}), opening intriguing perspectives for the study of spinor physics and impurities in ultradilute dipolar liquids.

%%%%%%%%%%%%%%%%%%%%%%%%%%%%%%%%%%%%%%%%%%%%%%%%%%%%%%%%%%%%%%%%%%%%%%%%%%%%%%%%%%%%%%%%%%%

% LHY ENERGY

\emph{LHY energy.--}  We first consider a homogeneous binary condensate of components $\sigma= a, b$, with densities $n_\sigma$, characterized by the intracomponent 
scattering lengths $a_{\sigma\sigma}$, the intercomponent scattering length $a_{ab}$, and the magnetic dipole moments $\mu_\sigma$~(our theory is equally valid for electric dipoles). 
All dipole moments are oriented by an external field along the same direction, $z$.  
For simplicity we consider equal masses $m_{a,b}=m$, although the formalism can be easily extended to unequal masses (for the experimentally relevant Er-Dy mixtures~\cite{Trautmann2018}, 
the masses are approximately equal). 

Using Hugenholz-Pines formalism~\cite{Hugenholz1959, footnote-SM}, we obtain the equation for the LHY energy density correction, $\epsilon_{LHY}$ \cite{footnote-Pastukhov}:
\begin{equation}
\!\!\epsilon_{LHY}(n_a,n_b)\!-\!\frac{1}{2}\sum_{\sigma}n_\sigma\frac{\partial}{\partial n_\sigma}\epsilon_{LHY}(n_a,n_b)\!=\!\chi(n_a,n_b),
\label{eq:HP}
\end{equation}
with
\begin{equation}
\chi(n_a,n_b) =  -\frac{1}{2} \int \frac{d^3k}{(2\pi)^3} \sum_{\lambda=\pm}
\frac{ [ \xi_\lambda(\vec k)-E(k)]^3}{4\xi_\lambda(\vec k)E(k)},
\end{equation}
where $\xi_\pm (\vec k)=[E(k)(E(k)+V_\pm(\theta_k))]^{1/2}$ are the Bogoliubov modes of the mixture,  
$E(k)=\hbar^2k^2/2m$,  and 
\begin{equation}
V_\pm (\theta_k)\!\!=\!\!\sum_{\sigma}\eta_{\sigma\sigma}n_\sigma \pm \sqrt{(\eta_{aa}n_a-\eta_{bb}n_b)^2+4\eta_{ab}^2n_an_b}.
\end{equation}
Above, $\theta_k$ is the angle between $\vec k$ and the dipole moments, 
$\eta_{\sigma\sigma'}(\cos\theta_k)=g_{\sigma\sigma'}+g_{\sigma\sigma'}^d (3\cos^2\theta_k-1)$, with 
$g_{\sigma\sigma'}=4\pi\hbar^2 a_{\sigma\sigma'}/m$, 
$g_{\sigma\sigma'}^d = \mu_0\mu_\sigma\mu_{\sigma'}/3=4\pi\hbar^2 a_{\sigma\sigma'}^{d}/m$, and $\mu_0$ is the vacuum permeability. 
The solution of Eq.~\eqref{eq:HP} is given by~\cite{footnote-Boudjemaa}:
\begin{equation}
\epsilon_{LHY} (n_a,n_b)\!=\!\! \frac{8}{15\sqrt{2\pi}}\left ( \frac{m}{4\pi\hbar^2}\right )^{\!\!\frac{3}{2}} \!\!\!\int \!\!d\theta_k \sin\theta_k \!\sum_{\lambda=\pm} V_\lambda(\theta_k)^{\!\frac{5}{2}}, 
\label{eq:LHY}
\end{equation}
which converges for 
$n_a=0$ or $n_b=0$ to the expression for a single-component dipolar BEC~\cite{Lima2011}, and for $\mu_{a,b}=0$ to 
that for a nondipolar mixture~\cite{Petrov2015}~(see~\cite{footnote-SM}).

From the form of $V_\pm(\theta_k)$ it is easy to see that $\epsilon_{LHY}=n^{5/2} F(P)$, where $n=n_a+n_b$ and $F$ is a 
function of the polarization $P=n_b/n_a$. A similar form occurs as well in nondipolar binary mixtures. However, for the latter, $P$ is homogeneously fixed at approximately $(g_{aa}/g_{bb})^{1/2}$ in the self-bound regime \cite{Petrov2015}. nondipolar self-bound mixtures are hence necessarily miscible, the LHY energy just depends on the total density, 
and the system is well approximated by 
an effective single-component model~\cite{Petrov2015}. In contrast, as discussed below, in a dipolar mixture the polarization is neither fixed nor homogeneous, 
resulting in rich spinor physics, including the possibility of immiscible droplets. 
 The problem is thus inherently a two-component one. In particular, the LHY energy is a function of the local densities of both components, and not only of the total density.

%%%%%%%%%%%%%%%%%%%%%%%%%%%%%%%%%%%%%%%%%%%%%%%%%%%%%%%%%%%%%%%%%%%%%%%%%%%%%%%%%%%%%%%%%%%

% FORMALISM

\emph{Formalism.--} We are interested in the ground state of self-bound dipolar mixtures. From Eq.~\eqref{eq:LHY}, we evaluate the LHY contribution 
to the chemical potentials, $\mu_{LHY}^{(\sigma)} (\{n_{a,b}\})= \partial\epsilon_{LHY}/\partial n_\sigma$.
As with single-component dipolar BECs~\cite{Waechtler2016} and nondipolar mixtures~\cite{Petrov2015}, we study spatially inhomogeneous dipolar mixtures by applying 
a local-density approximation (LDA) \cite{footnote-LDA} to the LHY term, $\mu_{LHY}^{(\sigma)}[\{n_{a,b}(\vec r)\}]$, obtaining two coupled Gross-Pitaevskii~(GP) equations which incorporate the 
effect of quantum fluctuations:
\begin{align}
i\hbar\frac{\partial}{\partial t}&\psi_\sigma(\vec r) = \Big [\frac{-\hbar^2\nabla^2}{2m} +  \sum_{\sigma'}\int d^3 r' V_{\sigma\sigma'}(\vec r-\vec r')n_{\sigma'}(\vec r') \nonumber \\
&+  \sum_{\sigma'} g_{\sigma\sigma'} n_{\sigma'}(\vec r) + \mu_{LHY}^{(\sigma)}[\{n_{a,b}(\vec r)\}] \Big ] \psi_\sigma(\vec r), 
\label{eq:GP}
\end{align}
where $n_\sigma(\vec r)\equiv|\psi_\sigma(\vec r)|^2$ and $V_{\sigma\sigma'}(\vec r)=\frac{\mu_0\mu_\sigma\mu_{\sigma'}}{4\pi r^3}(1-3\cos^2\theta)$, with $\theta$ the angle between $\vec r$ and the dipole moments. 

In addition to numerically intensive 3D simulations of Eqs.~\eqref{eq:GP}, we employ a simple variational approximation in the miscible regime using a Gaussian Anstatz:
\begin{equation}
\psi_{\sigma}(\vec r; l_\rho, l_z)=\left ( \frac{N_{\sigma}}{\pi^{3/2}l_{\rho}^{2}l_{z}} \right ) ^{1/2} e^{-\frac{1}{2}\left(\frac{\rho^{2}}{l_{\rho}^{2}}+\frac{z^{2}}{l_{z}^{2}}\right)}, 
\label{eq:Gaussian} 
\end{equation}
where $l_{\rho.z}$ are determined from energy minimization~\cite{footnote-SM}.  Ansatz~\eqref{eq:Gaussian} is, however, inappropriate for immiscible 
droplets~(see~\cite{footnote-SM} for an alternative ansatz in that regime).

%%%%%%%%%%%%%%%%%%%%%%%%%%%%%%%%%%%%%%%%%%%%%%%%%%%%%%%%%%%%%%%%%%%%%%%%%%%%%%%%%%%%%%%%%%%

% IMPURITY LIMIT

\emph{Impurity limit.--} The limit $N_b\ll N_a$ transparently illustrates the possible ground states of a dipolar mixture. The majority component is to a first approximation a single-component dipolar 
BEC, which remains self-bound for sufficiently large $N_a$ and low $a_{aa}/a_{aa}^d$~\cite{Schmitt2016, Baillie2016, Waechtler2016b}. 
Within the self-bound regime, the minority component experiences an effective 
potential induced by the majority component:
\begin{equation}
\!\!\mu_{ab}(\vec r) \simeq g_{ab} n_a(\vec r)  +\! \int \! d^3 r' V_{ab}(\vec r\!-\!\vec r')n_a(\vec r') + \gamma_{ab} n_a(\vec r)^{\frac{3}{2}},
\label{eq:Eff_Pot}
\end{equation}
where $\gamma_{ab}=\frac{32}{3\sqrt{\pi}}\left ( \frac{m}{4\pi\hbar^2} \right )^{\frac{3}{2}}\int_0^1 du\, \eta_{aa}(u)^{\frac{1}{2}} \eta_{ab}(u)^2$. 
The last term in Eq.~\eqref{eq:Eff_Pot} is the 
zero-momentum beyond-mean-field correction of the polaron energy resulting from the interaction of the impurity with the elementary excitations of the majority component. 
This repulsive term is crucial for the miscibility of the mixture.
It favors immiscibility, reducing the critical $a_{ab}$ by tens of $a_0$.
Take the example of $N_a=1270$, $N_b \to 0$, and $a_{aa}=70a_0$. When $\gamma_{ab}$ is properly included we find that immiscibility occurs at $a_{ab}\simeq 75a_0$, whereas excluding $\gamma_{ab}$ pushes the immiscibility threshold up to $a_{ab}\simeq 115a_0$.

\begin{figure} 
\begin{center}
\includegraphics[width=1\columnwidth]{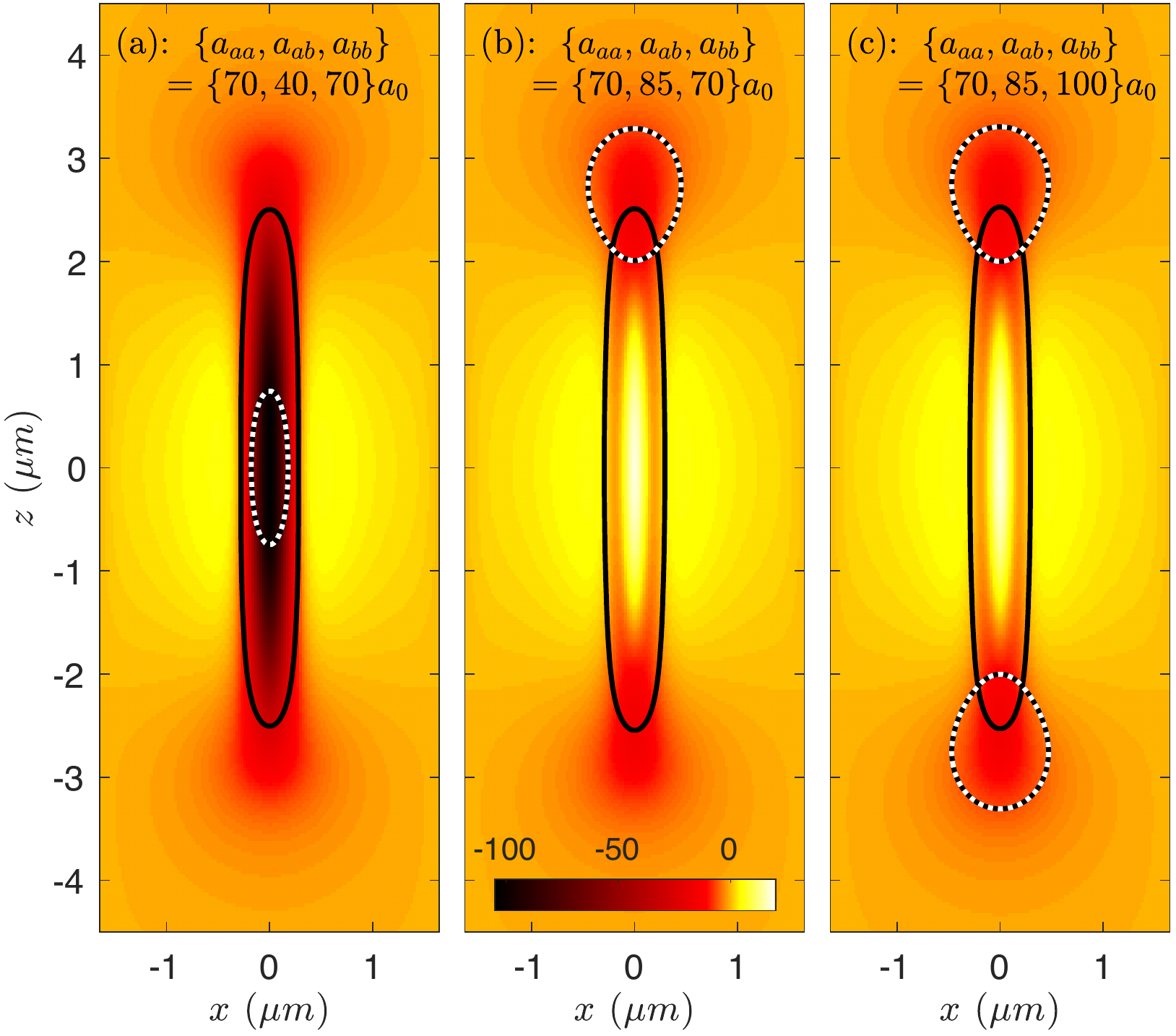}
\caption{Effective potential $\mu_{ab}(x,y=0,z)$ [arb.~unit] experienced in the impurity limit
by the minority component in (a) miscible, (b) asymmetric immiscible, and (c) symmetric immiscible regimes. The majority component ($N_a = 2000$) is represented by a black density contour, while the impurity component ($N_b=20$) contour is white-black dotted -- both are drawn at 10\% of the respective peak densities.
 \label{fig:2}}
\end{center}
\end{figure}

Dipolar attraction dominates for small-enough $g_{ab} > 0$, resulting in a minimum of  $\mu_{ab}(\vec r)$ at the droplet center, see Fig.~\ref{fig:2}(a). 
Component $b$ then remains within the droplet and the mixture is miscible. In contrast, for large-enough $g_{ab}$, $\mu_{ab}(\vec r)$ develops a maximum 
at the droplet center~(Figs.~\ref{fig:2}(b,c)). In the absence of dipolar interactions the minority component would be ejected. However, crucially, 
the partially attractive and long-range character of the dipolar interaction results in two potential minima, along the dipole direction, $z$, which extend outside the $a$ droplet~(Figs.~\ref{fig:2}(b,c)).
With increasing $g_{ab}$, component $b$ is pushed away from the droplet center, first developing two $\mu_{ab}(\vec r)$ minima while still miscible, 
and is eventually positioned outside component $a$ in complete immiscibility.
A sufficiently large $g_{bb}>0$ favors an  equal occupation of both minima~(Fig.~\ref{fig:2}(c)), whereas for smaller $g_{bb}$ the $b$ component will be biased towards one of the minima, spontaneously breaking the discrete 
$Z_2$ symmetry~(Fig.~\ref{fig:2}(b)). As shown below, although the energy scales interplay differently for more balanced populations,   
the same three self-bound ground states still occur: miscible, symmetric immiscible and asymmetric immiscible.

%%%%%%%%%%%%%%%%%%%%%%%%%%%%%%%%% 

% FIGURE 2

%%%%%%%%%%%%%%%%%%%%%%%%%%%%%%%%% 

%%%%%%%%%%

% FIGURE 3

\begin{figure} 
\begin{center}
\includegraphics[width=1\columnwidth]{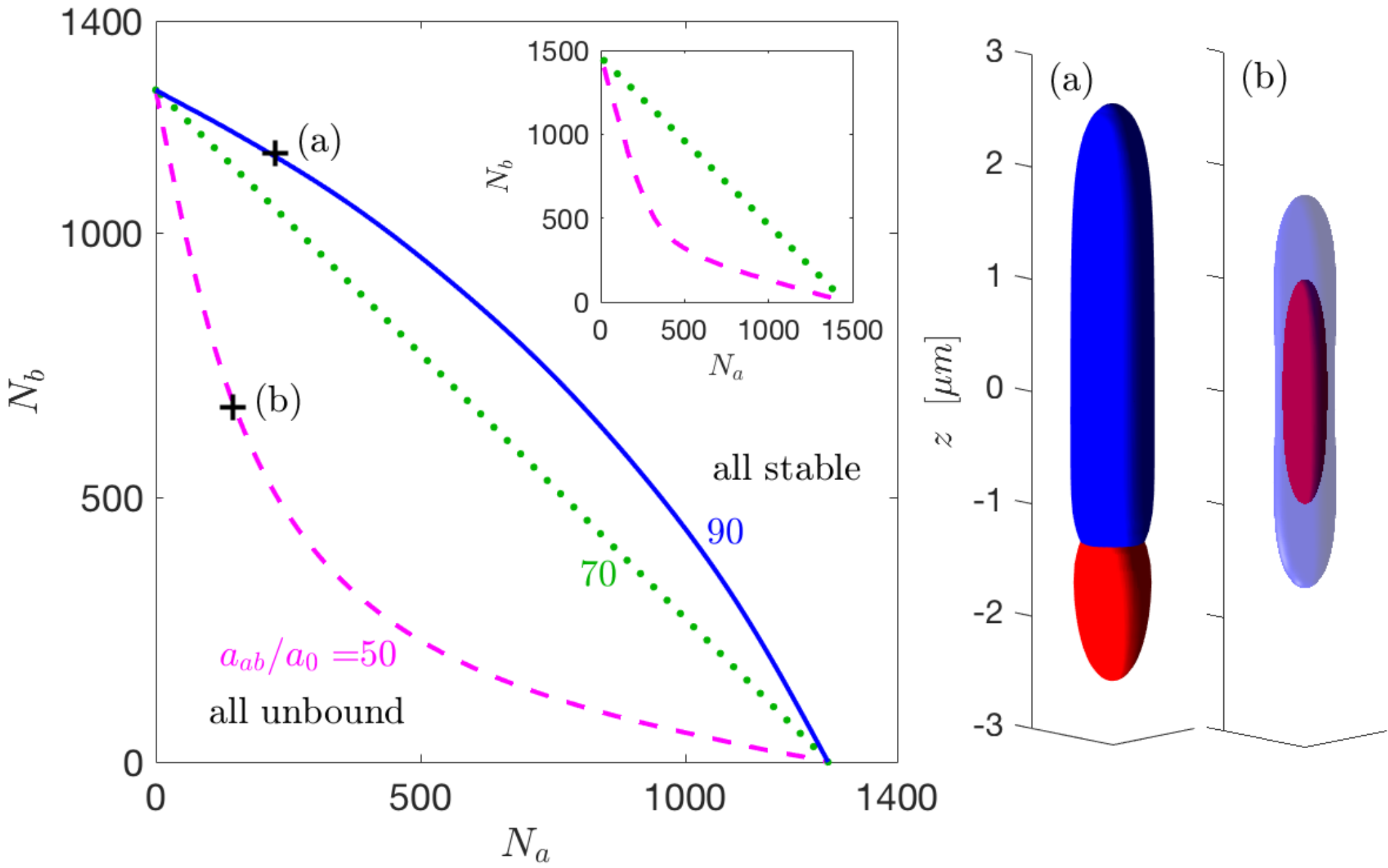}
\caption{Instability threshold as a function of particle number in each component for a Dy-Dy mixture with $a_{aa}=a_{bb}=70a_0$, and 
$a_{ab}=50a_0$, $70a_0$ and $90a_0$. The mixture remains self-bound for the parameter region above the curves. 
The inset shows the results obtained using the variational ansatz~\eqref{eq:Gaussian} for $a_{ab}=50a_0$ and $70a_0$.
The subplots show the 3D probability contour 
for the $a$~(red) and $b$~(blue) component, drawn at $10\%$ of the respective peak densities. 
 \label{fig:1}}
\end{center}
\end{figure}

%%%%%%%%%%

%%%%%%%%%%%%%%%%%%%%%%%%%%%%%%%%%%%%%%%%%%%%%%%%%%%%%%%%%%%%%%%%%%%%%%%%%%%%%%%%%%%%%%%%%%%

% NUMERICAL RESULTS

\emph{Self-bound miscible and immiscible droplets.--}  Figures~\ref{fig:3} and~\ref{fig:1} summarize our GP results of the ground-state physics for a 
Dy-Dy mixture~($a_{aa,bb}^{d}=129.2 a_0$, with $a_0$ the Bohr radius), for equal intracomponent interactions, $a_{aa,bb}=70a_0$. 
Figure~\ref{fig:3} shows the phase diagram for a fully balanced mixture~($N_{a,b}=N/2$), as a function of the total particle number $N$ and $a_{ab}$. 
The self-bound--unbound transition is marked by a solid curve. Within the self-bound regime, the system experiences an abrupt phase transition (dotted line) from a miscible regime at low $a_{ab}$~[see Fig.~\ref{fig:3}(a)] to an asymmetric immiscible regime for large $a_{ab}$~[Fig.~\ref{fig:3}(b)]. 
For the particular case of Figs.~\ref{fig:3} and~\ref{fig:1}, where the intracomponent interactions and the dipole moments are equal,  
the miscible-immiscible threshold is approximately independent of the number of atoms. In more general cases, as illustrated below, the transition may be driven by changing the 
particle number.

While in the impurity limit the droplet stability only depends on the intracomponent interactions of the majority component, independently of the miscibility or immiscibility of the 
mixture, in the balanced case there is a marked interplay between droplet stability and miscibility. 
When decreasing $a_{ab}$ into the miscible regime, the droplet becomes significantly more stable. In particular, the critical total number of particles 
for self-binding falls considerably, see Fig.~\ref{fig:3}. The dashed line in the figure shows the instability boundary predicted by the Gaussian ansatz~\eqref{eq:Gaussian}, 
which reproduces well the GP results within the miscible regime.

The instability threshold presents a marked dependence on the polarization $N_a/N_b$ of the mixture. In Fig.~\ref{fig:1}, we depict the stability threshold as a function of $N_a$ and $N_b$, for 
$a_{ab}=50a_0$, $70a_0$ and $90a_0$ for the same case as Fig.~\ref{fig:3}. In the impurity limit, as mentioned above, the stability does not depend on $a_{ab}$ and all curves converge to the critical 
number for a single component. Deep within the miscible regime~($a_{ab}=50a_0$), balanced droplets have a much lower critical total number, $N_{cr}$, for stability compared to the single-component case. For $a_{ab}=50a_0$, $N_{cr}\simeq  700$ for $N_a=N_b$, i.e. just $350$ particles in each component, whereas $N_{cr}\simeq 1270$ for $N_{a}=0$ or $N_b=0$, showing that the mutual confinement strongly 
reinforces self-binding. 

%%%%%%%%%%%%%%%%%%%%%%%%%%%%%%%%% 

% FIGURE 4

\begin{figure} 
\begin{center}
\includegraphics[width=1\columnwidth]{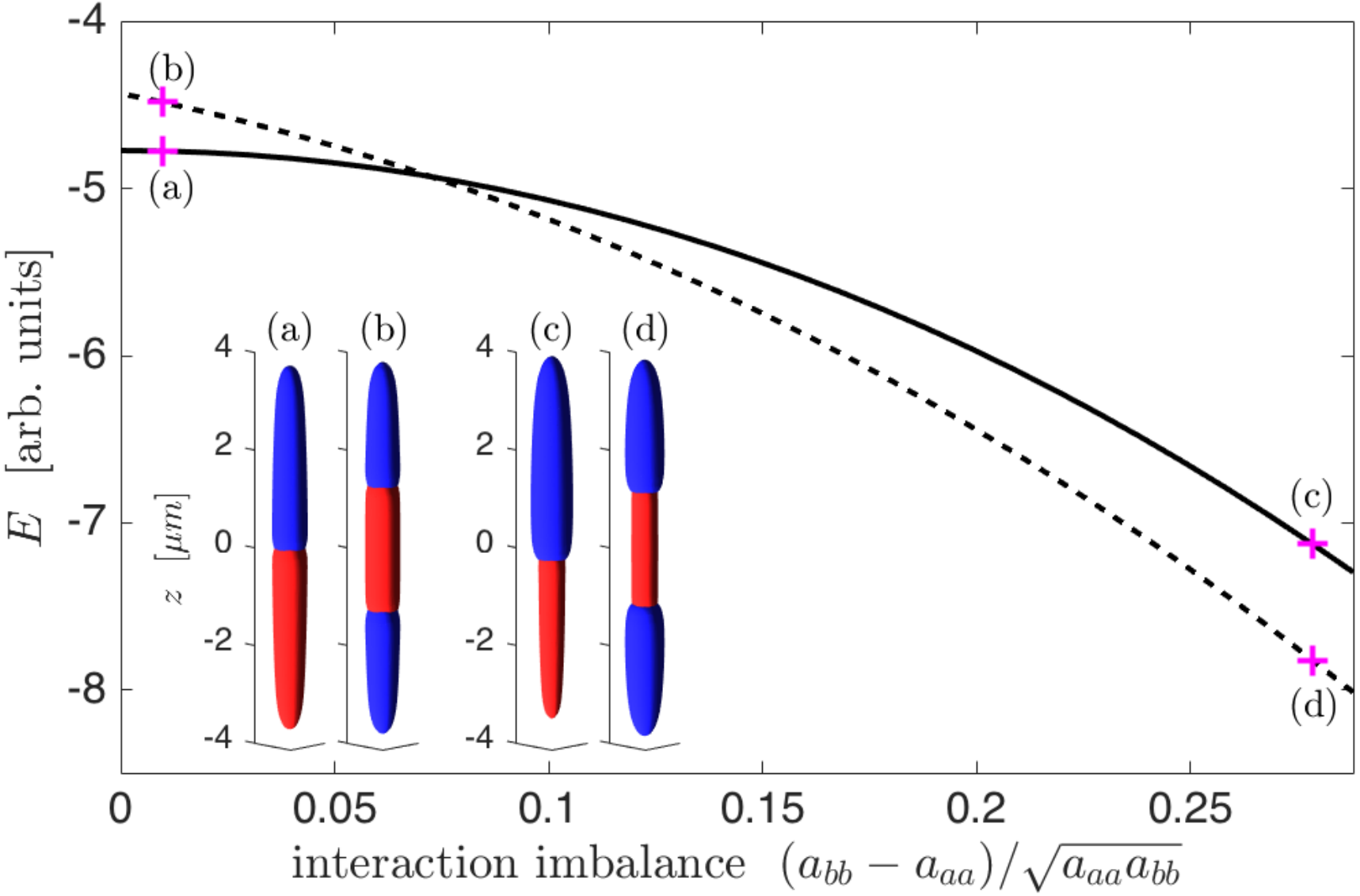}
\caption{Asymmetric immiscible-to-symmetric immiscible transition.  
Energy~ of the symmetric~(dashed) and asymmetric~(solid) immiscible phase as a function of $(a_{bb}-a_{aa})$ for $a_{ab}=85a_0$, and $N_{a,b}=2000$. The subplots show 3D contours for the $a$~(red) and $b$~(blue) components, drawn at $5\%$ of the respective peak densities.}
 \label{fig:4}
\end{center}
\end{figure}

%%%%%%%%%%%%%%%%%%%%%%%%%%%%%%%%% 

In the immiscible regime, a {\em droplet molecule} forms, i.e.~a self-bound solution of two attached droplets. 
The repulsion resulting from the intercomponent mean-field contact term and the LHY energy~\cite{footnote-LHY-separation}
results in phase separation. For the particular cases in Figs.~\ref{fig:3} and~\ref{fig:1}, this separation is always asymmetric, see Fig.~\ref{fig:3}(b) and Fig.~\ref{fig:1}(a)~(the 
latter should be compared to Fig.~\ref{fig:2}(b) in the impurity limit).  In more general scenarios, as illustrated below, the interplay between intra- and intercomponent interactions may favor 
a symmetric configuration with two domain walls~(as in Fig.~\ref{fig:2}(c) in the impurity limit). In any case, as in the impurity limit, the droplets remain attached despite their phase separation due to the 
intercomponent dipole-dipole interactions. Each component creates at its borders (and beyond) an attractive potential pocket in which the other component is trapped, leading to mutual attachment. 
The attractive interactions exerted by the other component lead not only to attachment, but also to reinforced stability. 
As shown in Fig.~\ref{fig:1}, for the immiscible regime~($a_{ab}=90a_0$), in contrast to the miscible case, $N_{cr}$ grows when the mixture is more balanced~($N_{cr}\simeq 1500$ for $N_{a}=N_b$). 
Even so, only $N_{a,b}=750$ particles in each component are necessary for self-binding -- compared to $\simeq 1270$ in the single-component case -- showing that despite phase separation, the mutual attachment allows for the stabilization of two droplets that would be individually unstable. 
The instability threshold flattens within the immiscible regime~(Fig.~\ref{fig:3}), due to the drastic reduction of the intercomponent overlapping, but  
the non-negligible dependence on $a_{ab}$ shows that the width of the domain wall remains finite.

%%%%%%%%%%%%%%%%%%%%%%%%%%%%%%%%% 

% FIGURE 5
    
\begin{figure} 
\begin{center}
\includegraphics[width=\columnwidth]{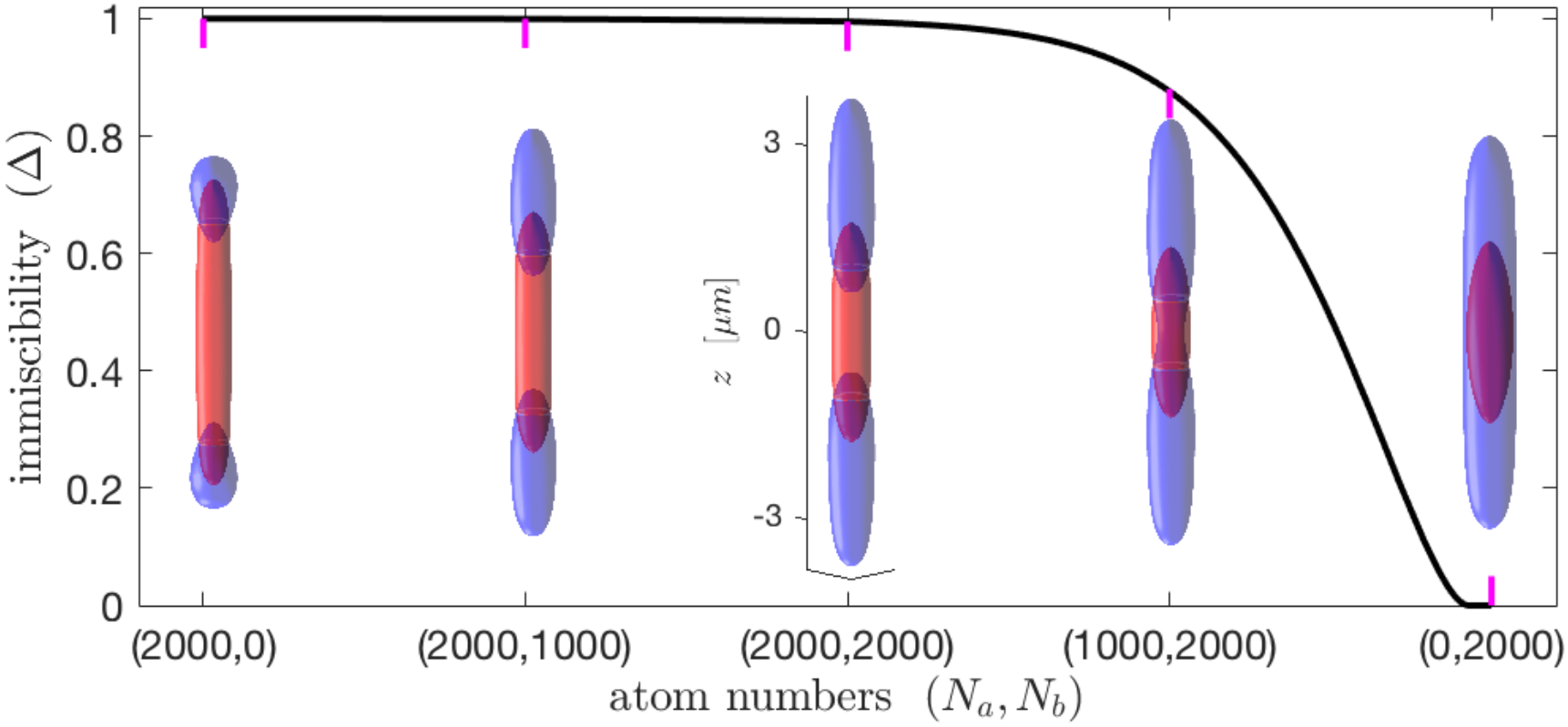}
\caption{Symmetric immiscible-to-miscible crossover. Contrast $\Delta$ (see text) for different $(N_a,N_b)$ going from $(2000,0)$ to $(2000,2000)$, and then from $(2000,2000)$ to $(0,2000)$, for imbalanced interactions $(a_{aa},a_{ab},a_{bb})/a_0 = (65,70,75)$. Subplots show the 3D density contours for $a$~(red) 
and $b$~(blue) components for the impurity limits, $(1000,2000)$, $(2000,2000)$, and $(2000,1000)$. All contours are drawn at $5\%$ of the respective peak densities.}
 \label{fig:5}
\end{center}
\end{figure}

%%%%%%%%%%%%%%%%%%%%%%%%%%%%%%%%% 

While the cases considered above display a miscible-to-asymmetric immiscible transition, an immiscible-immiscible transition may also occur between a symmetric and asymmetric configuration, as illustrated in 
Fig.~\ref{fig:4}, where we consider $N_a=N_b=2000$, $a_{ab}=85a_0$, and $(a_{aa}+a_{bb})/2=70a_0$. This figure shows that the population distribution may be changed not only by modifying $a_{ab}$ but also by changing the ratio $a_{aa}/a_{bb}$.
While for $a_{aa}=a_{bb}$ the asymmetric configuration has a lower energy compared to the symmetric one, the symmetric configuration becomes the ground-state at a critical $a_{bb}-a_{aa}$, marking the onset of a first order phase transition.
The symmetric immiscible solution can be the ground state -- overcoming the cost of two domain walls -- because the component with the smaller intraspecies contact interactions forms a narrower droplet (see Fig.~\ref{fig:4} insets). Not only does this reduce its internal dipolar energy, it also creates deeper attractive potential pockets at both ends, within which the second component can equally divide itself to minimize energy.

Finally, the symmetric immiscible configuration may crossover to a miscible phase, as illustrated in Fig.~\ref{fig:5}, where we consider $(a_{aa},a_{ab},a_{bb})/a_0 = (65,70,75)$.
We monitor the crossover by considering the contrast, $\Delta\equiv |n_{a0}/n_{am}-n_{b0}/n_{bm}|$, where 
$n_{\sigma m}$ is the maximal density of component $\sigma$, and $n_{\sigma 0}$ is its density at the droplet center~\cite{Lee2016}. The system undergoes a symmetric immiscible-to-miscible crossover when $N_b/N_a$ grows.
This is because in the impurity limit, $N_b \to 0$,  $(a_{aa}, a_{ab})/a_0 =  (65, 70)$ leads to an immiscible mixture [ca.~Fig. 2(c)], whereas  for $N_a \to 0$,  $(a_{ab}, a_{bb})/a_0 = (70, 75)$ results in miscibility. Note that component $a$ always remains at the center since $a_{aa}$ is the lowest.
Furthermore, we should point out that more generally all possible transitions discussed in this paper can occur as a function of the polarization.
This opens the possibility of an intriguing scenario. In typical mixture experiments, three-body losses are larger in one of the two components~\cite{Cabrera2018,Semeghini2018}. While for nondipolar mixtures losses in one component leads to the unraveling of the whole self-bound mixture~\cite{Cabrera2018,Semeghini2018}, in dipolar mixtures losses may instead result in a loss-induced 
miscible-immiscible crossover or transition.

%%%%%%%%%%%%%%%%%%%%%%%%%%%%%%%%%%%%%%%%%%%%%%%%%%%%%%%%%%%%%%%%%%%%%%%%%%%%%%%%%%%%%%%%%%%

% CONCLUSIONS

\emph{Conclusions.--} While nondipolar Bose mixtures are necessarily miscible with approximately fixed polarization, dipolar Bose mixtures present a rich array of spinor physics, and in particular 
may undergo a miscible-immiscible transition. We have shown that self-bound mixtures may be in three different ground states: a miscible droplet, and immiscible droplet "molecules" -- in either a symmetric or asymmetric configuration -- and we illustrated the different phase transitions and crossovers between these phases. We also discussed the impurity limit, in which beyond mean-field corrections of the polaron energy play a crucial role in the miscibility of the mixture.  
Dipolar mixtures free the spinor physics of self-bound ultradilute liquids, opening exciting perspectives for the study of ultracold superfluid-superfluid mixtures -- exhibiting similar physics to that of $^4$He-$^3$He 
droplets and much more, including: the dynamics of superfluid-superfluid droplets~(e.g.~under rotation~\cite{Pi2020}); probing superfluidity of one component by another; polaron physics 
in low-dimensional dipolar mixtures~\cite{Ardila2019}; loss-induced miscibility transitions; Bose-Fermi droplets; and supersolid-supersolid mixtures.

%%%%%%%%%%%%%%%%%%%%%%%%%%%%%%%%%%%%%%%%%%%%%%%%%%%%%%%%%%%%%%%%%%%%%%%%%%%%%%%%%%%%%%%%%%%

% ACKNOWLEDGMENTS

\begin{acknowledgments}
We thank L.~Chomaz and F.~Ferlaino for stimulating discussions.
We acknowledge support of the Deutsche Forschungsgemeinschaft (DFG, German Research Foundation) under Germany's Excellence Strategy -- EXC-2123 QuantumFrontiers -- 390837967, 
and FOR 2247. RNB acknowledges the European Union's Horizon 2020 research and innovation programme under the Marie Sk{\l}odowska-Curie grant agreement No.~793504 (DDQF).

Note added: After the completion of this work we became aware of a related work \cite{Smith2020}, whose results are compatible and complementary to ours.
\end{acknowledgments}

%%%%%%%%%%%%%%%%%%%%%%%%%%%%%%%%%%%%%%%%%%%%%%%%%%%%%%%%%%%%%%%%%%%%%%%%%%%%%%%%%%%%%%%%%%%

% BIBLIOGRAPHY

%% SM
\pagebreak
\onecolumngrid
\begin{center}
\textbf{\Large Supplementary Material}
\end{center}
\beginsupplement

\section{Derivation of the LHY correction}
We briefly discuss further details on the derivation of the LHY correction of Eq.~(4) of the main text. 
The Hugenholz-Pines~(HP) formalism may be easily extended to two-component condensates. As discussed in the main text, the 
LHY energy density, $\epsilon_{LHY}$ obeys the differential equation:
\begin{equation}
\!\!\epsilon_{LHY}(n_a,n_b)\!-\!\frac{1}{2}\sum_{\sigma}n_\sigma\frac{\partial}{\partial n_\sigma}\epsilon_{LHY}(n_a,n_b)\!=\!\chi(n_a,n_b).
\label{eq:HP2}
\end{equation}
$\chi(n_a,n_b)$, which is given by Eq. (2) of the main text, can be re-written in the form:
$\chi(n_a,n_b)= \frac{\hbar^2}{m} (n_a a_{aa})^{5/2} G(P)$, where $G(P)$ is a function of the polarization $P=n_b/n_a$. 
We employ then the ansatz $\epsilon_{LHY}=\frac{\hbar^2}{m} (n_a a_{aa})^{5/2} \tilde G(P)$. Note that 
$\sum_\sigma n_\sigma \frac{\partial}{\partial n_\sigma}P = 0$, and hence 
$\sum_\sigma n_\sigma \frac{\partial}{\partial n_\sigma} \epsilon_{LHY} = \frac{5}{2} \epsilon_{LHY}$. As a result, the HP equation is greatly simplified: $\epsilon_{LHY}(n_a,n_b) = -4\chi(n_a,n_b)$, and hence 
\begin{eqnarray}
\epsilon_{LHY}(n_a,n_b)&=&2\int \frac{d^3k}{(2\pi)^3} \sum_\nu \frac{(\xi_\nu(k,\theta_k)-E(k))^3}{4\xi_\nu(k,\theta_k)E(k)} 
\nonumber \\ 
&=& \left ( \frac{2m}{\hbar^2} \right )^{3/2}\frac{1}{8\pi^2}\int_0^\pi d\theta_k \sin\theta_k  \sum_{\lambda=\pm} V_\lambda(\theta_k)^{5/2} \int_0^\infty dq q^2 \frac{\left (\sqrt{q^2+1}-q \right )^3}{\sqrt{q^2+1}}
\nonumber \\ 
&=& \frac{8}{15\sqrt{2\pi}}\left ( \frac{m}{4\pi\hbar^2}\right )^{\frac{3}{2}} \int d\theta_k \sin\theta_k \sum_{\lambda=\pm} V_\lambda(\theta_k)^{\frac{5}{2}},
\label{eq:SM-LHY}
 \end{eqnarray}
as in Eq.~(4) of the main text. For a single-component dipolar BEC~($n_b= 0$, $a_{aa}=a$, $a_{aa}^d/a=\epsilon_{dd}$), Eq.~\eqref{eq:SM-LHY} becomes of the form:
\begin{equation}
\frac{E_{LHY}}{V}=  \frac{64}{15} g n^2 \left ( \frac{n a^3}{\pi}\right )^{1/2} \int_0^1 du (1+\epsilon_{dd}(3u^2-1))^{5/2},
\end{equation}
recovering the result of Ref.~\cite{Lima2011}. For nondipolar binary mixtures~($a_{aa}^d=a_{bb}^d=0$), Eq.~\eqref{eq:SM-LHY} becomes 
\begin{equation}
\frac{E_{LHY}}{V}=  \frac{8}{15\pi^2} \left ( \frac{m}{\hbar^2} \right )^{3/2} (g_{aa} n_a)^{5/2} f\left ( \frac{a_{ab}^2}{a_{aa}a_{bb}},\frac{a_{bb}n_b}{a_{aa}n_a} \right ),
\end{equation}
with $f(x,y)=\frac{1}{4\sqrt{2}}\sum_{\sigma=\pm} \left (1+y\pm\sqrt{(1-y)^2+4xy} \right )^{5/2}$, recovering the result of Ref.~\cite{Petrov2015}.

%%%%%%%%%%%%%%%%%%%%%%%%%%%%%%%%%%%%%%%%%%%%%%%%%%%%%%%%%%%%%%%%%%

\section{Variational calculations} 
 
\subsection{Gaussian ansatz} 

Assuming miscibility of the mixture, we may consider a Gaussian ansatz,  $n_{\sigma}(\vec{\mathbf{r}};l_{\rho},l_{z})=\left|\psi(\vec{\mathbf{r}},l_{\rho},l_{z})\right|^{2}$ (see Eq.~(6) of the main text),
\begin{equation}
n_{\sigma}(\vec{\mathbf{r}};l_{\rho},l_{z})=\frac{N_{\sigma}}{\pi^{3/2}l_{\rho}^{2}l_{z}}e^{-\left(\frac{\rho^{2}}{l_{\rho}^{2}}+\frac{z^{2}}{l_{z}^{2}}\right)}.
\end{equation}
Using this ansatz we may evaluate the total energy as a function of the variational widths $l_{\rho,z}$: \begin{eqnarray}
E[l_\rho, 
l_z]&=&\frac{\hbar^2}{4m}\left[\frac{2}{l_{\rho}^{2}}+\frac{1}{l_{z}^{2}}\right] \sum_\sigma N_\sigma \nonumber 
\\
&+& \frac{1}{2(2\pi)^{3/2} l_\rho^2l_z} \sum_{\sigma,\sigma'} N_\sigma 
N_{\sigma'} \left [g_{\sigma\sigma'}+g^{d}_{\sigma\sigma'} f(\kappa) 
\right ]  \nonumber \\
&+& \frac{32}{75\sqrt{5\pi}}\!\! \left ( 
\frac{m}{4\pi^{5/2}\hbar^2l_\rho^2l_z} \right )^{\!\!\frac{3}{2}} \!\!\!
\sum_{\lambda=\pm} \int_0^1 \!\! Q_\lambda(u)^{\!\frac{5}{2}}du,
\end{eqnarray}
with $g_{\sigma\sigma'}$ and $g^{d}_{\sigma\sigma'}$ the contact and dipolar coupling strengths defined in the main text. In addition,  $f(\kappa)=\frac{2\kappa^2+1}{\kappa^2-1}-\frac{3\kappa^2 
\arctan\sqrt{\kappa^2-1}}{(\kappa^2-1)^{3/2}}$ with $\kappa=\frac{l_\rho}{l_z}$ the aspect ratio, and $Q_\pm(u)=\sum_\sigma N_\sigma \eta_{\sigma\sigma}(u)\pm
\sqrt{(N_a\eta_{aa}(u)-N_b\eta_{bb}(u))^2+4\eta_{ab}(u)^2N_aN_b}$, and the functions $\eta_{\sigma\sigma'}$ are defined in the main text.\\

%%%%%%%%%%%%%%%%%%%%%%

\subsection{Flat-top ansatz} The Gaussian ansatz discussed previously is not suitable for treating immiscible mixtures. 
For this purpose we employ an alternative ansatz, where 
we assume that the density profile of the droplet is Gaussian radially and flat-top axially:
%(see wavefunction Eq. [7]).

%%%%%%%%%%%%%%%%%%%%

\begin{equation}
n_{\sigma}(\vec{\mathbf{r}};L_{\rho},L_{\sigma})=\frac{N_{\sigma}}{\pi L_{\rho}^{2}L_{\sigma}}e^{-\left(\frac{\rho}{L_{\rho}}\right)^{2}}\Pi\left(\frac{z+z_{\sigma}}{L_{\sigma}}\right),
\end{equation}
where $\Pi(x)=1$ if $|x|<1/2$ and zero otherwise. Note that in this ansatz, we allow for different axial lengths $L_\sigma$ -- where $\sigma=\left\{ a,b\right\}$ -- and center-of-mass (COM) positions of the components, $z_\sigma$. Miscibility with an axial flat-top density profile is  captured by this ansatz when $z_{a,b}=0$. Energy is minimized with respect to four variational parameters: $L_\rho$, $L_a$, $L_b$, and the displacement $\Delta z_{\sigma,\sigma'}=\left|z_{\sigma}-z_{\sigma}'\right|$. 
The energy as a function of the variational parameters is of the form:
\begin{eqnarray}
E[L_\rho, 
L_\sigma]&=&\frac{\hbar^2}{2mL_{\rho}^2}\sum_\sigma N_\sigma \nonumber 
\\
&+& \frac{1}{4\pi L_{\rho}^{2}}\sum_{\sigma,\sigma'}\frac{N_{\sigma}N_{\sigma'}}{\sqrt{L_{\sigma}L_{\sigma'}}}\left[g_{\sigma\sigma'}\Lambda_{\sigma\sigma'}^{c}+g_{\sigma\sigma'}^{d}\Lambda_{\sigma\sigma'}^{d}\right] \nonumber \\
&+& \frac{4}{75\pi^{2}\sqrt{2}}\left(\frac{m}{\pi\hbar^{2}L_{\rho}^{2}\sqrt{L_{a}L_{b}}}\right)^{3/2}\int dz\int d\theta_{k}\sin\theta_{k}\sum_{\lambda=\pm}\left[S_{\lambda}(\cos\theta_{k})\right]^{5/2}.
\end{eqnarray}

Here we employ the auxiliary functions 
\begin{eqnarray}
\Lambda_{\sigma\sigma'}^{c}&=&\frac{1}{2}\left(\sqrt{\frac{L_{\sigma}}{L_{\sigma'}}}+\sqrt{\frac{L_{\sigma'}}{L_{\sigma}}}\right)-\frac{\Delta z_{\sigma,\sigma'}}{\sqrt{L_{\sigma}L_{\sigma'}}}\\
\Lambda_{\sigma\sigma'}^{d}&=&\frac{1}{2\pi}\int dk_{z}h_{\sigma\sigma'}(k_{z})\mathrm{sinc}\left(\frac{k_{z}L_{\sigma}}{2}\right)\mathrm{sinc}\left(\frac{k_{z}L_{\sigma'}}{2}\right)\exp\left(-ik_{z}\frac{\Delta z_{\sigma,\sigma'}}{\sqrt{L_{\sigma}L_{\sigma'}}}\right),
\end{eqnarray}
where $h_{\sigma\sigma'}(k_{z})=\int dk_{\rho}k_{\rho}\left[\frac{3k_{z}^{2}}{(k_{\rho}/\kappa_{\sigma\sigma'})^{2}+k_{z}^{2}}-1\right]\exp\left(-\frac{1}{2}k_{\rho}^{2}\right)=-1-3ue^{u}\mathrm{Ei}\left (- \left(\frac{k_{z}\kappa_{\sigma\sigma'}}{\sqrt{2}}\right)^{2} \right )$, with $\kappa_{\sigma\sigma'}=L_{\rho}/\sqrt{L_{\sigma}L_{\sigma'}}$, 
and $S_{\pm}(\cos\theta_{k})=\sum\eta_{\sigma\sigma}n_{\sigma}^{z}\pm\sqrt{\left(\eta_{aa}n_{a}^{z}-\eta_{bb}n_{b}^{z}\right)^{2}+4\eta_{ab}^{2}n_{a}^{z}n_{b}^{z}}$, with  $n_{a}^{z}=\sqrt{\frac{L_{b}}{L_{a}}}\Pi\left(\frac{z+z_{a}}{L_{a}}\right)$ and $n_{b}^{z}=\sqrt{\frac{L_{a}}{L_{b}}}\Pi\left(\frac{z+z_{b}}{L_{b}}\right)$.

%%%%%%%%%%%%%%%%%%%%%%

% FIGURE SM1

 \begin{figure} 
\begin{center}
\includegraphics[width=3in]{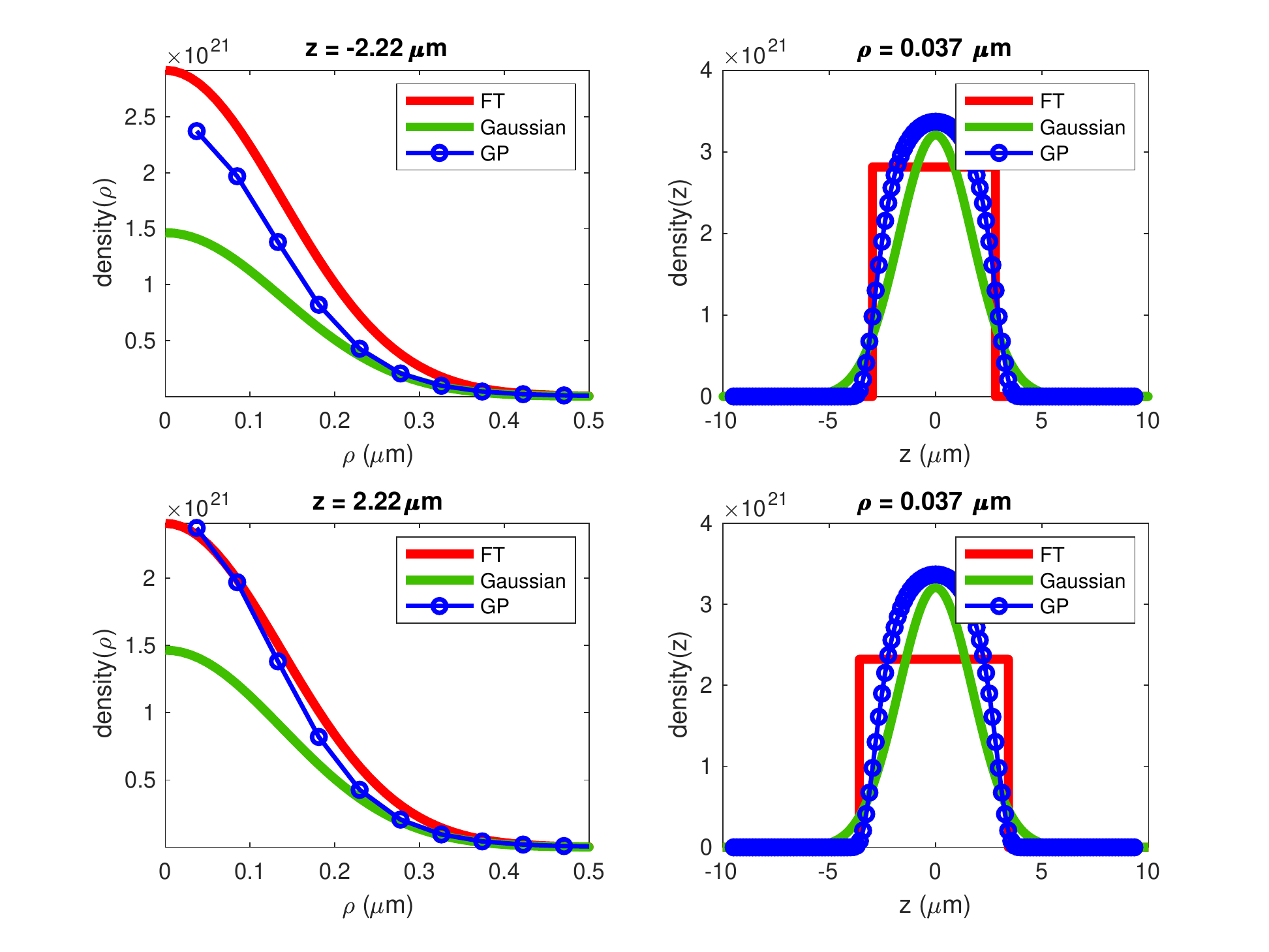}
\includegraphics[width=3in]{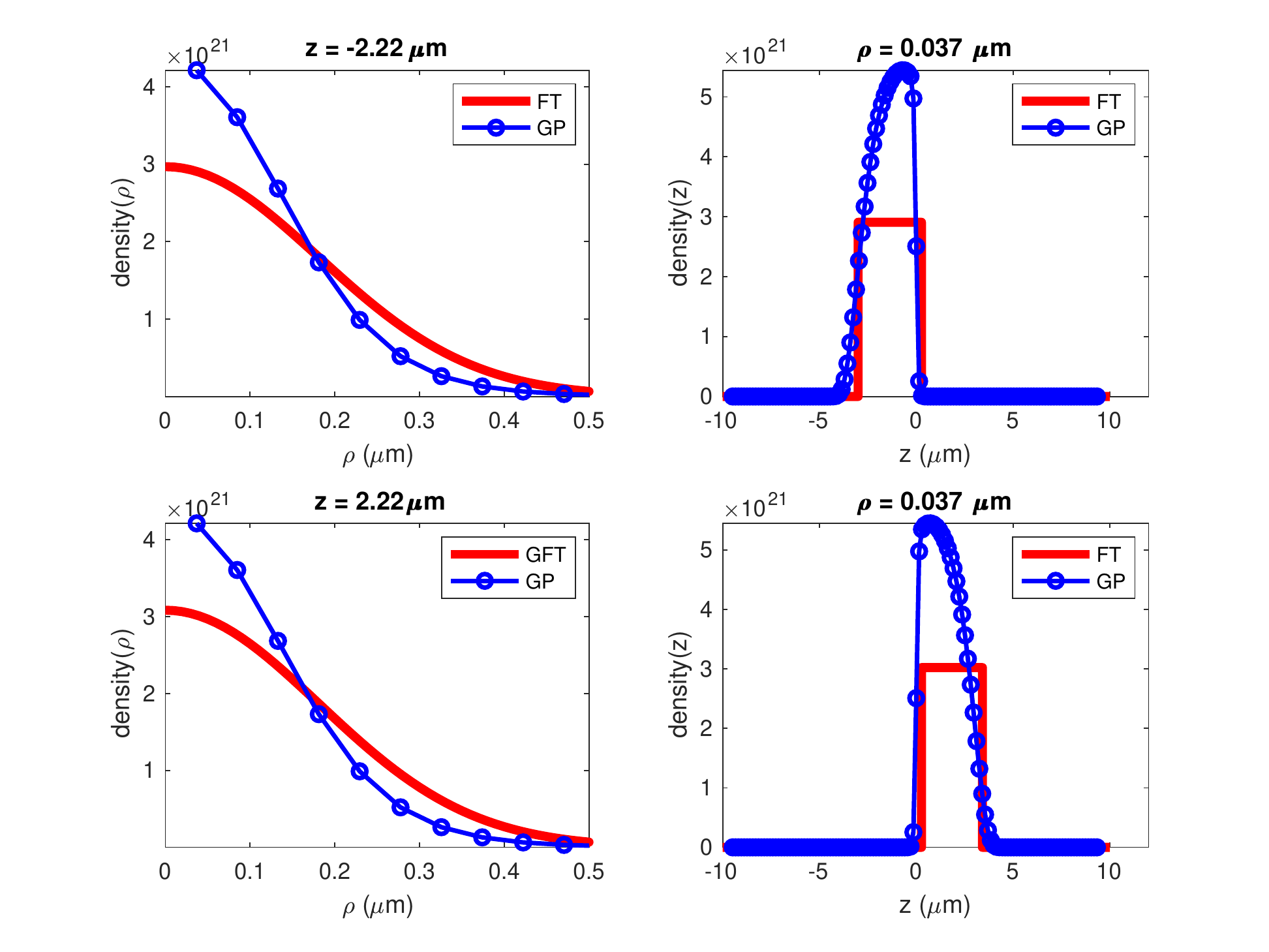}
\caption{Radial and axial density profiles for the case, $(N_{a},N_{b})=(2000,2000)$ and $a_{aa}=a_{bb}=70a_0$. The (left) miscible regime $a_{ab}=64.5a_0$ and (right) immiscible regime $a_{ab}=85a_0$. We compare the results of the fully-Gaussian ansatz, the flat-top~(FT) ansatz, and the GP calculations.
}
 \label{Fig1S}
\end{center}
\end{figure}

%%%%%%%%%%%%%%%%%%%%%%

%%%%%%%%%%%%%%%%%%%%%%%%%%%%%%

% FIGURE SM2

\begin{figure} 
\begin{center}
\includegraphics[width=3in]{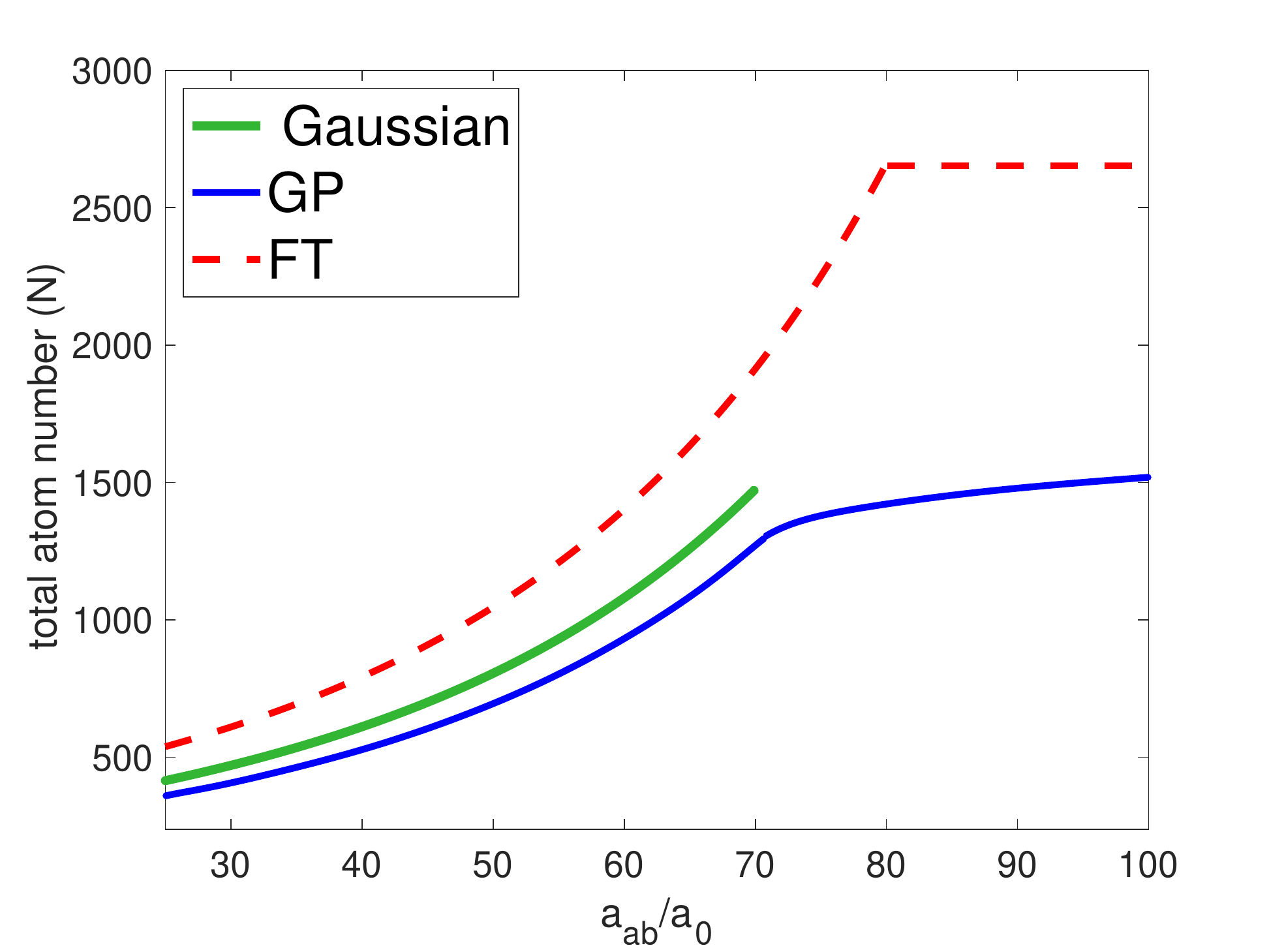}
\caption{Critical number of particles for stability of the fully balanced ($N_a=N_b$) self-bound solution for $a_{aa}=a_{bb}=70a_0$. We compare the results of the fully-Gaussian ansatz, the flat-top~(FT) ansatz, and the GP calculations.}
\label{Fig2S}
\end{center}
\end{figure}

%%%%%%%%%%%%%%%%%%%%%%%%%%%%%%

%%%%%%%%%%%%%%%%%%%%%%

\subsection{Variational results} 

In Fig.~\ref{Fig1S}, we plot the radial and axial density profiles for $N_{a,b}=2000$ and $a_{ab}=64.5a_0$~(miscible regime) and $a_{ab}=85a_0$~(strongly-immiscible regime). In the former case, both the fully-Gaussian ansatz and the flat-top ansatz are compared against the GP prediction. Qualitatively, both ansatzes give a good description of the radial density profile, but the fully-Gaussian density profile gives a better quantitative agreement, especially at the self-bound/unbound transition (not explicitly shown, but see the good agreement with the full GP solutions for the stability boundary in Fig.~\ref{Fig2S}).

For the strongly immiscible case, the fully Gaussian ansatz is no longer adequate, since the axially displaced Gaussians do not provide a good description of the the domain-wall region, which is typically characterized by a sharply-changing density (see the GP results in Fig.~\ref{Fig1S}, right).
%For the immiscible case,  the fully Gaussian ansatz is no longer %adequate. Moreover an ansatz with two axially displaced Gaussians %does not provide a good approximation, since when the center-of-mass %of the components do not coincide that ansatz greatly misrepresents %the domain-wall region, which is crucial for the description of %immiscibility.
In contrast, the flat-top ansatz captures well the qualitative features of the domain-wall region. Note, however, that the flat-top ansatz is only suitable for the asymmetric immiscible case.

As shown in Fig.~\ref{Fig2S}, the flat-top ansatz qualitatively reproduces the miscible/immiscible transition (see the kink at $a_{ab}^{cr}\simeq 80a_0$), which is moderately shifted compared to the GP result ($a_{ab}^{cr}\simeq 70a_0$).
%The ansatz provides a transition at $a_{ab}^{cr}=80a_0$, which should %be compared with the GP result, $a_{ab}^{cr}=70a_0$.
For the flat-top ansatz, the immiscible solution is always fully immiscible, and the critical number of particles remains constant for $a_{ab}>a_{ab}^{cr}$. However, the flat-top ansatz significantly overestimates the critical number of particles in the immiscible regime, by close to a factor of $2$.


\begin{thebibliography}{99}

\bibitem{Toennies2001} J. P. Toennies, A. F. Vilesov, and K. B. Whaley , Physics Today 54, 2, 31 (2001). 

\bibitem{Toennies2004} J. P. Toennies and A. F. Vilesov, Angew. Chem. Int. Ed. {\bf 43}, 2622 (2004). 

\bibitem{Barranco2006} M. Barranco, R. Guardiola, E. S. Hern\'andez, R. Mayol, J. Navarro, and M. Pi, J. Low Temp. Phys. {\bf 142}, 1 (2006).

\bibitem{Ancilotto2017} F. Ancilotto, M. Barranco, F. Coppens, J. Eloranta, N. Halberstadt, A. Hernando, D. Mateo, and M. Pi., Int. Rev. Phys. Chem. {\bf 36}, 621 (2017).

\bibitem{Grebenev1998} S. Grebenev, J.P. Toennies, and A.F. Vilesov, Science {\bf 279}, 2083 (1998).

\bibitem{Harms1999} J. Harms, M. Hartmann, B. Sartakov, J. P. Toennies, and A. F. Vilesov, J. Chem.  Phys. {\bf 110}, 5124 (1999). 

\bibitem{Pi2020} M. Pi, F. Ancilotto, J. M. Escart\'in, R. Mayol, and M. Barranco, Phys. Rev. B {\bf 102} 060502(R) (2002).

\bibitem{Stienkemeier2006} F. Stienkemeier and K. K. Lehman, J. Phys. B: At. Mol. Opt. Phys. {\bf 39}, R127 (2006). 

\bibitem{Choi2006} M. Y. Choi, G. E. Douberly, T. M. Falconer, W. K. Lewis, C. M. Lindsay, J. M. Merrit, P. L.  Stiles, and R. E. Miller, Int. Rev. Phys. Chem. {\bf 25}, 15 (2006). 

\bibitem{Tiggesbaeumker2007} J. Tiggesb\"aumker and F. Stienkemeier, Phys. Chem. Chem. Phys. {\bf 9}, 4748 (2007).

\bibitem{Szalewicz2008} K. Szalewicz, Int. Rev. Phys. Chem. {\bf 27}, 273 (2008). 

\bibitem{Kadau2016} H. Kadau, M. Schmitt, M. Wenzel, C. Wink, T. Maier, I. Ferrier-Barbut, and T. Pfau, Nature (London) {\bf 530}, 194 (2016).

\bibitem{Chomaz2016} L. Chomaz, S. Baier, D. Petter, M. J. Mark, F. W\"achtler, L. Santos, and F. Ferlaino, Phys. Rev. X {\bf 6}, 041039 (2016).

\bibitem{Schmitt2016} M. Schmitt, M. Wenzel, F. B\"ottcher, I. Ferrier-Barbut, and T. Pfau, Nature (London) {\bf 539}, 259 (2016).

\bibitem{Cabrera2018} C. R. Cabrera, L. Tanzi, J. Sanz, , B. Naylor, P. Thomas, P. Cheiney, and L. Tarruell, Science {\bf 359}, 301 (2018).

\bibitem{Semeghini2018} G. Semeghini, G. Ferioli, L. Masi, C. Mazzinghi, L. Wolswijk, F. Minardi, M. Modugno, G. Modugno, M. Inguscio, and M. Fattori, Phys. Rev. Lett. {\bf 120}, 235301 (2018).

\bibitem{DErrico2019} C. D'Errico, A. Burchianti, M. Prevedelli, L. Salasnich, F. Ancilotto, M. Modugno, F. Minardi, and C. Fort, Phys. Rev. Research {\bf 1}, 033155 (2019).

\bibitem{Petrov2015} D. S. Petrov, Phys. Rev. Lett. {\bf 115}, 155302 (2015).

\bibitem{Waechtler2016} F. W\"achtler and L. Santos, Phys. Rev. A {\bf 93}, 061603(R) (2016).

\bibitem{Rakshit2019} D. Rakshit, T. Karpiuk, M. Brewczyk, and M. Gajda, SciPost Phys. {\bf 6}, 079 (2019).

\bibitem{Trautmann2018} A. Trautmann, P. Ilzh\"ofer, G. Durastante, C. Politi, M. Sohmen, M. J. Mark,  and F. Ferlaino, Phys. Rev. Lett. {\bf 121}, 213601 (2018).

\bibitem{Durastante2020} G. Durastante, C. Politi, M. Sohmen, P. Ilzh\"ofer, M. J. Mark, M. A. Norcia, and F. Ferlaino, Phys. Rev. A, {\bf 102}, 033330 (2020).

\bibitem{Wenzel2018} M. Wenzel, T. Pfau, and I. Ferrier-Barbut,  Phys. Scr. {\bf 93}, 104004 (2018).

\bibitem{footnote-atoms} We take the atomic mass for both components to be 161.9u and the dipole moments as 9.93$\mu_{\bf B}$.

\bibitem{Hugenholz1959} N. M. Hugenholtz and D. Pines, Phys. Rev. {\bf 116}, 489 (1959).

\bibitem{footnote-SM} See the Supplementary Material for more details concerning the derivation of the LHY correction and the variational calculations.

\bibitem{footnote-Pastukhov} See \cite{Pastukhov2017} for a discussion 
of other effects of quantum fluctuations in dipolar Bose mixtures.

\bibitem{Pastukhov2017} V. Pastukhov, Phys. Rev. A {\bf 95}, 023614 (2017).

\bibitem{footnote-Boudjemaa} See Ref.~\cite{Boudjemaa2018} for an alternative derivation, which results in an implicit form of the LHY energy correction for a homogeneous 3D dipolar Bose mixture. The formalism we employ provides the explicit expression~\eqref{eq:LHY} and does not require us to cure divergences. The latter makes our formalism better suited to treat lower- and cross-dimensional problems~\cite{Edler2017,Igl2019}.

\bibitem{Boudjemaa2018} A. Boudjemaa, Phys. Rev. A {\bf 98}, 033612 (2018).

\bibitem{Edler2017} D. Edler, C. Mishra, F. W\"achtler, R. Nath, S. Sinha, and L. Santos, Phys. Rev. Lett. {\bf 119}, 050403 (2017).

\bibitem{Igl2019}  T. Ilg, J. Kumlin, L. Santos, D. S. Petrov, and H. P. B\"uchler, Phys. Rev. A {\bf 98}, 051604(R) (2018).

\bibitem{Lima2011} A. R. P. Lima and A. Pelster, Phys. Rev. A {\bf 84}, 041604(R) (2011).

\bibitem{footnote-LDA} The LDA may be locally compromised when the droplet density steeply changes, such as for very sharp domain walls. As with single-component dipolar BECs and nondipolar Bose mixtures we expect that corrections to the LDA should at most lead to quantitative deviations of the boundaries between the different phases. The overall qualitative picture should  be preserved.

\bibitem{Baillie2016} D. Baillie, R. M. Wilson, R. N. Bisset, and P. B. Blakie, Phys. Rev. A {\bf 94}, 021602(R) (2016).

\bibitem{Waechtler2016b} F. W\"achtler and L. Santos, Phys. Rev. A {\bf 94}, 043618 (2016).

\bibitem{footnote-LHY-separation} Note, however, that in contrast to the mean-field terms, it is not possible to separate the intra- and intercomponent contributions to the LHY term.

\bibitem{Lee2016} K. L. Lee, N. B. J\o rgensen, I.-K. Liu, L. Wacker, J. J. Arlt, and N. P. Proukakis, Phys. Rev. A {\bf 94}, 013602 (2016).

\bibitem{Ardila2019} L. A. Pe\~na Ardila and T. Pohl, 
J. Phys. B: At. Mol. Opt. Phys. {\bf 52}, 015004 (2019).

\bibitem{Smith2020} J. C. Smith, D. Baillie, P. B. Blakie, Phys. Rev. Lett. {\bf 126}, 025302 (2021).

\end{thebibliography}
\end{document}